\documentclass[twocolumn,showpacs,aps,prl,superscriptaddress]{revtex4}

\usepackage{graphicx}
\usepackage{dcolumn}
\usepackage{amsmath}
\usepackage{epsfig}

\input pubboard/babarsym
\newcommand{\beq}{\begin{equation}}
\newcommand{\eeq}{\end{equation}}
\newcommand{\bea}{\begin{eqnarray}}
\newcommand{\eea}{\end{eqnarray}}

\newcommand{\DE}{\ensuremath{\Delta E}}

\newcommand{\pvec}{{\bf p}}

\def\Y#1S{{\Upsilon\rm(#1S)}}
\def\Ks{{K^0_{\scriptscriptstyle S}}}

\newcommand{\BABARPubYear}    {06}
\newcommand{\BABARPubNumber}  {043}

\newcommand{\SLACPubNumber} {11965}

\def\figurebox#1#2#3{%
    \def\arg{#3}%
    \ifx\arg\empty
    {\hfill\vbox{\hsize#2\hrule\hbox to #2{\vrule\hfill\vbox to #1{\hsize#2\vfill}\vrule}\hrule}\hfill}%
    \else
    {\hfill\epsfbox{#3}\hfill}%
    \fi}

\begin{document}

\preprint{\babar-PUB-\BABARPubYear/\BABARPubNumber} 
\preprint{SLAC-PUB-\SLACPubNumber} 

\begin{flushleft}
\babar-PUB-\BABARPubYear/\BABARPubNumber\\
SLAC-PUB-\SLACPubNumber\\
\end{flushleft}

\title{
{\large \bf \boldmath
Branching Fraction Measurements of Charged \B\ Decays \\ to $K^{*+}K^+K^-$, $K^{*+}\pi^+K^-$, $K^{*+}K^+\pi^-$ and $K^{*+}\pi^+\pi^-$  Final States} 
}

%
\author{B.~Aubert}
\author{R.~Barate}
\author{M.~Bona}
\author{D.~Boutigny}
\author{F.~Couderc}
\author{Y.~Karyotakis}
\author{J.~P.~Lees}
\author{V.~Poireau}
\author{V.~Tisserand}
\author{A.~Zghiche}
\affiliation{Laboratoire de Physique des Particules, F-74941 Annecy-le-Vieux, France }
% ...
\collaboration{The \babar\ Collaboration}
\noaffiliation

\date{\today}

\begin{abstract}
Branching fraction and asymmetry measurements of charmless $B^+\rightarrow K^{*+}h^+_1h^-_2$ (where $h_{1,2}$ = $K$, $\pi$) decays are presented, using a data sample of 232 million $\Upsilon(4S) \rightarrow$ \BB\ decays collected with the \babar\ detector at the SLAC PEP-II asymmetric-energy $B$ factory. Using a maximum likelihood fit, the following branching fraction results were obtained: ${\cal B}$($B^+ \rightarrow K^{*+}K^+ K^-)$ = (36.2 $\pm$ 3.3 $\pm$ 3.6) $\times$ 10$^{-6}$ and ${\cal B}$($B^+$ $\rightarrow$ $K^{*+}\pi^+\pi^-$) = (75.3 $\pm$ 6.0 $\pm$ 8.1)  $\times$ 10$^{-6}$. Upper limits were set for ${\cal B}$($B^+$ $\rightarrow$ $K^{*+}\pi^+ K^-$) $<$ 11.8 $\times$ 10$^{-6}$ and ${\cal B}$($B^+$ $\rightarrow$ $K^{*+}K^+ \pi^-$) $<$ 6.1 $\times$ 10$^{-6}$ at 90\% confidence level. The charge asymmetries for the decays $B^+ \rightarrow K^{*+}K^+ K^-$ and $B^+$ $\rightarrow$ $K^{*+}\pi^+\pi^-$ were measured to be  ${\cal A}_{K^*KK} = 0.11 \pm 0.08 \pm 0.03$ and ${\cal A}_{K^*\pi\pi} = 0.07 \pm 0.07 \pm 0.04$,  respectively. The first error quoted on branching fraction and asymmetry measurements is statistical and the second systematic. 
\end{abstract}


\maketitle
\begin{figure}[htb]
\resizebox{\columnwidth}{!}{
\includegraphics{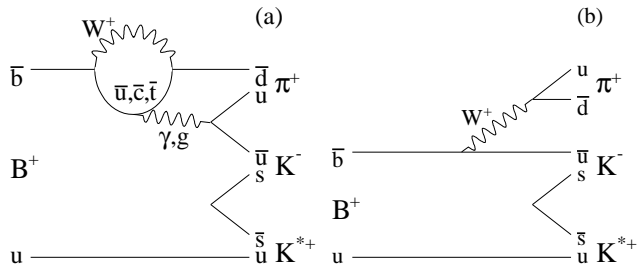}
}
\caption{Penguin (a) and tree (b) Feynman diagrams for the decay $B^+$ $\rightarrow$ $K^{*+}\pi^+ K^-$.
}\label{fig:feynKstkpi}
\end{figure}

 Charmless decays of $B$ mesons to three-body final states are very important in aiding the understanding of the weak interaction and complex quark couplings described by the Cabibbo-Kobayashi-Maskawa (CKM) matrix elements~\cite{ckm}. Improved experimental measurements of these charmless decays, combined with theoretical developments, can provide significant constraints on the CKM matrix parameters or uncover evidence for physics beyond the Standard Model. For example, the branching fraction of the decay $B^+$ $\rightarrow$ $K^{*+}\pi^+ K^-$ is sensitive to CKM matrix elements $V_{td}$ and $V_{ub}$ (see Figure~\ref{fig:feynKstkpi}). Additionally, a $B^+$ $\rightarrow$ $K^{*+}\pi^+ K^-$ branching fraction value equal to or smaller than the branching fraction of the Standard Model suppressed decay $B^+$ $\rightarrow$ $K^{*+}K^+\pi^-$ would be an indication of new physics.

We identify $K^{*+}$ mesons through their decay to $\Ks\pi^+$. Charged $B$-meson decays to $\Ks\pi^+h^+_1 h^-_2$  (where $h_{1,2}$ = $K$ or $\pi$) are dominated by $K^{*+}h^+_1 h^-_2$, but can also proceed via a nonresonant component as well as through observed intermediate charmless resonances such as $B^+ \rightarrow K^{*+}\phi$ or $B^+ \rightarrow K^{*+}\rho^0$~\cite{babarkstphi,babarkstrho,bellekstphi}, or other as-yet-unobserved intermediate charmless resonances. To date, there are only limits on the charmless decays $B^+ \rightarrow K^{*+}h^+_1 h^-_2$, measured by the ARGUS experiment~\cite{argus1991} using less than 0.2 fb$^{-1}$.

Asymmetry measurements of charmless $B$ decays can be used to probe for $C\!P$ violation where the $C\!P$ asymmetry is defined as:

\begin{equation}
{\cal A}_{K^*h_1h_2} = \frac{\Gamma_{K^{*-}h^-_1h^+_2} - \Gamma_{K^{*+}h^+_1h^-_2}}{\Gamma_{K^{*-}h^-_1h^+_2} + \Gamma_{K^{*+}h^+_1h^-_2}} ~,
\end{equation}

\noindent and $\Gamma$ is the partial width of the charged $B$ decay in the subscript.  

In the analyses presented in this paper, branching fractions of $B^+ \rightarrow K^{*+}K^+ K^-$ and $B^+ \rightarrow K^{*+}\pi^+ \pi^-$  were measured for the first time, and upper limits were set for $B^+ \rightarrow K^{*+}\pi^+ K^-$ and the Standard Model suppressed decay $B^+ \rightarrow K^{*+}K^+ \pi^-$, where charge-conjugate decays are also implied. The selection criteria required events with a reconstructed $\Ks\pi^+ h^+_1 h^-_2$ final state such that the total charmless contribution to the $K^{*+}h^+_1h^-_2$  Dalitz plot could be measured (with charmed and charmonium resonances removed), including contributions from resonant charmless substructure. Finally, the ${\cal A}_{K^*h_1h_2}$ values for the observed decays $B^+ \rightarrow K^{*+}K^+ K^-$ and $B^+ \rightarrow K^{*+}\pi^+ \pi^-$ were measured.

The data used in this analysis were collected at the \pep2\ asymmetric-energy \epem\ storage ring with the \babar\ detector~\cite{babar}. The \babar\ detector consists of a double-sided five-layer silicon tracker, a 40-layer drift chamber, a Cherenkov detector, an electromagnetic calorimeter and a magnet with instrumented flux return. The data sample has an integrated luminosity of 210~fb$^{-1}$ collected at the $\FourS$ resonance, which corresponds to $(231.8\pm 2.5)\times 10^6$ \BB\ pairs. It was assumed that the $\FourS$ decayed equally to neutral and charged $B$-meson pairs. In addition, 21.6 fb$^{-1}$ of data collected at 40~MeV below the $\FourS$ resonance were used for background studies.

Candidate $B$ mesons were reconstructed from three tracks and a $\Ks$, where the $\Ks$ was reconstructed from $\pi^+\pi^-$ candidates.  Each of the three tracks that were not associated with the $\Ks$ were required to have at least 12 hits in the drift chamber, a transverse momentum greater than 100~\mevc and to be consistent with originating from the beam-spot.  These tracks were identified as either pion or kaon candidates using energy loss (\dedx) measured in the tracking system and the number of photons measured by the Cherenkov detector and their corresponding Cherenkov angles. Furthermore, the tracks were required to fail the electron selection based on \dedx information, their ratio of energy in the calorimeter to momentum in the drift chamber, and the shower shape of the signal in the calorimeter. The $\Ks$ candidates were required to have a reconstructed mass within 15~MeV/$c^2$ of the nominal $K^0$ mass~\cite{pdg}, a decay vertex separated from the $B^+$ decay vertex by at least five standard deviations, and a cosine of the angle between the line joining the $B$ and $\Ks$ decay vertices and the $\Ks$ momentum greater than 0.999.

To characterize signal events, three kinematic variables and one event-shape variable were used. The first kinematic variable $\DeltaE$, is the difference between the center-of-mass (CM) energy of the $B$-candidate and $\sqrt{s}/2$, where $\sqrt{s}$ is the total CM energy. The second is the beam-energy-substituted mass $\mes = \sqrt{(s/2 + \pvec_i \cdot \pvec_B)^2/E_i^2 - \pvec^2_B}$, where  $\pvec_B$ is the $B$ momentum and  ($E_i, \pvec_i$) is the four-momentum of the $\FourS$ in the laboratory frame.  The third kinematic variable is the $\Ks \pi^+$ invariant mass, $m_{K^*}$, used to identify $K^{*+}$ candidates. Using these three kinematic variables, candidates were required to be in the ranges  $|\Delta E| <0.1 \gev$, $5.25<\mes<5.29 \gevcc$ and 0.772 $<$ $m_{K^*}$ $<$ 0.992 GeV/$c^2$. The event-shape variable is a Fisher discriminant ($\mathcal{F}$)~\cite{Fisher}, constructed from a linear combination of the cosine of the angle between the $B$-candidate momentum and the beam axis, the cosine of the angle between the $B$-candidate daughters thrust axis and the beam axis, and the zeroth and second angular moments of energy flow about the thrust axis of the reconstructed $B$.

Continuum quark production ($e^+e^-$ $\rightarrow$ $q\bar{q}$ where $q$ = {\em u,d,s,c}) was the dominant source of background. This was suppressed using another event-shape variable which was the cosine of the angle $\theta_T$ between the thrust axis 
of the selected $B$-candidate and the thrust axis of the rest of the event. For continuum background, the distribution $|\cos\theta_T|$ is strongly 
peaked towards unity whereas the distribution is flat for signal events. Therefore, the relative amount of continuum background was reduced by requiring $|\cos\theta_T| < 0.8$.

Simulated Monte Carlo (MC) events were used to study background from other $B$-meson decays. The largest $B$-background for $B^+ \rightarrow K^{*+}\pi^+ \pi^-$ candidates comes from decays including charmonium mesons such as $J/\psi K^{*+}$, $\chi_{c0}K^{*+}$ and $\psi(2S)K^{*+}$, where the charmonium meson decays to $\mu^+\mu^-$ which are misidentified as pions, or where the charmonium meson decays directly to $\pi^+\pi^-$. These background events were removed by vetoing reconstructed $\pi^+\pi^-$ masses in the range 3.04 $<$ $m_{\pi^+\pi^-}$ $<$ 3.17~GeV/$c^2$, 3.32 $<$ $m_{\pi^+\pi^-}$ $<$ 3.53~GeV/$c^2$ and 3.60 $<$ $m_{\pi^+\pi^-}$ $<$ 3.78~GeV/$c^2$, identifying the $J/\psi$, $\chi_{c0}$ and  $\psi(2S)$ mesons, respectively. For $B^+ \rightarrow K^{*+}K^+ K^-$ candidates, $J/\psi K^{+}$ and $\chi_{c0}K^{*+}$ events were removed by rejecting events with a reconstructed invariant mass in the range 3.04 $<$ $m_{K^{*+}K^-}$ $<$ 3.17~GeV/$c^2$ and 3.32 $<$ $m_{K^+K^-}$ $<$ 3.53~GeV/$c^2$,  respectively. 

Potential charm contributions from $B^+$ $\rightarrow$ $\bar{D^0}(\rightarrow K^{*+}h^-_1)h^+_2$ events were removed from corresponding $B^+ \rightarrow K^{*+}h^+_2 h^-_1$ candidates by vetoing events with a  reconstructed $K^{*+}h^-_1$ invariant mass in the range 1.83 $<$ $m_{K^*h}$ $<$ 1.91~GeV/$c^2$. Additional decays such as $B^+$ $\rightarrow$ $D^-(\rightarrow \Ks\pi^-)\pi^+\pi^+$ and $B^+ \rightarrow K^{*+}\bar{D^0}(\rightarrow K^+ \pi^-)$ were removed from $B^+ \rightarrow K^{*+}\pi^+ \pi^-$ and $B^+ \rightarrow K^{*+}\pi^+ K^- /K^{*+}K^+ \pi^-$ candidates, respectively by vetoing events with reconstructed $K\pi$ invariant masses in the range 1.83 $<$ $m_{K\pi}$ $<$ 1.91~GeV/$c^2$, using the same veto range for $\bar{D^0}$ and $D^+$ candidates. Studies of MC events showed that the largest remaining charmed background was  $B^+$ $\rightarrow$ $\bar{D^0}(\rightarrow K^{*+}\pi^-)\pi^+$, with 18\% of these events passing the veto. Surviving charmed events had a reconstructed $D$ mass outside the veto as a result of using the wrong $\pi^+$, $K^+$ or $\Ks$ which was incorrectly selected from the other $B$ decay in the event. 

After the above selection criteria were applied, a fraction of events for all decays had more than one candidate. For those events, one candidate alone was selected by choosing the candidate whose fitted $B$ decay vertex had the smallest $\chi^2$ value. Studies of MC events showed the selection of $B^+ \rightarrow K^{*+}\pi^+ \pi^-$ events produced the largest number of multiple candidates, in 29\% of events, where for these multiple candidates the correct one was reconstructed 70\% of the time.

After all requirements, there were five main sources of $B$-background: two-body decays proceeding via a charmonium meson; two and three-body decays proceeding via a $D$ meson; combinatorial background from three unrelated particles ($K^{*+}h^+_1 h^-_2$);  charmless two or four-body $B$ decays with an extra or missing particle and three-body decays with one or more particles misidentified. Along with selection efficiencies obtained from MC simulation, existing branching fractions for these modes \cite{hfag} were used to estimate their background contributions which were included in fits to data. 

In order to extract the signal event yield for the channel under study, an unbinned extended maximum likelihood fit was used. The likelihood function for $N$ candidates is:

\begin{equation}
  \label{eq:Likelihood}
  \mbox{$\mathcal{L}$} \,=\, \frac{1}{N!}\exp\left(-\sum_{i=1}^{M} n_i\right)\, \prod_{j=1}^N 
\,\left(\sum_{i=1}^M n_{i} \, P_{i}(\vec{\alpha},\vec{x_j})\right) ~,
\end{equation}

\noindent where $i$ and $j$ are integers, $M$ is the number of hypotheses (signal, continuum background and $B$-background), $n_i$ is the number of events for each hypothesis determined by maximizing the likelihood function and $P_{i}(\vec{\alpha},\vec{x_j})$ is a probability density function (PDF) with the parameters $\vec{\alpha}$ associated with $\vec{x}$, where $\vec{x}$ can be any of the four variables \mes, \DE, $\mathcal{F}$, and $m_{K^*}$. The PDF is a product $P_{i}(\vec{\alpha},\vec{x}) = P_{i}(\vec{\alpha}_{\mes} ,\mes ) \cdot P_{i}(\vec{\alpha}_{\DE} ,\DE ) \cdot  P_{i}(\vec{\alpha}_{\mathcal{F}}, \mathcal{F}) \cdot  P_{i}(\vec{\alpha}_{m_{K^*}}, {m_{K^*}})$. Studies of MC simulations showed correlations between these variables were small for signal and continuum background hypotheses. However for $B$-background, correlations were observed between \mes and \DE, which were taken into account by forming a 2-dimensional PDF for these variables. The parameters of the signal and $B$-background PDFs were determined from MC simulation. The continuum background parameters were allowed to vary in the fit, to help reduce systematic effects from this dominant event type. Upper sideband data, defined to be in the region $0.1 < \Delta E <0.3 \gev$ and $5.25<\mes<5.29 \gevcc$, was used to model the continuum background PDFs. For the \mes\ PDFs, a Gaussian distribution was used for signal, a threshold function~\cite{argus} for continuum and the combination of a Gaussian and threshold function for $B$-background. For the \DE\ PDFs, a sum of two Gaussian distributions with the same mean was used for the signal, a first-order polynomial for the continuum background and the combination of the two Gaussians and a first-order polynomial was used for $B$-background. The $\cal{F}$ signal, continuum and $B$-background PDFs were described using the sum of two Gaussian distributions with distinct means and widths. Finally for $m_{K^*}$ PDFs, the sum of a Breit-Wigner and a first-order polynomial was used to describe the signal, continuum and $B$-background distributions. The first-order polynomial component of the $m_{K^*}$ PDFs was used to model misreconstructed events for signal and background. 

In order to allow uncertainties and corrections due to MC simulation to be calculated and applied to the signal modes under study, the decay $B^+ \rightarrow \bar{D^0}(\rightarrow K^{*+} \pi^-) \pi^+$ was used as a calibration channel. These events were selected using the $B^+ \rightarrow K^{*+}\pi^+ \pi^-$ selection criteria, but requiring the reconstructed $K^{*+}\pi^-$ invariant mass be in the range $1.84<m_{K^{*+}\pi^-}<1.88 \gevcc$.  With more than 1800 signal events and approximately a 4 to 1 signal to background ratio in the total number of $B^+ \rightarrow \bar{D^0}(\rightarrow K^{*+} \pi^-) \pi^+$ candidates, it was possible to fit the signal PDF parameters for this mode. 

Branching fractions, $\mathcal{B}$, are usually calculated  using the following equation, $\mathcal{B}$\,=\ $n_{sig}/(N_{\myBB} \times \epsilon$), where $n_{sig}$ is the fitted number of signal events, $\epsilon$ is the average signal efficiency obtained from MC simulation and $N_{\myBB}$ is the total number of \BB\ events. For the charmless $B^+ \rightarrow K^{*+}h^+_1 h^-_2$  branching fraction, the average efficiency cannot be taken directly from MC events. This was because the efficiency varies over the Dalitz plane and the distribution of events in the Dalitz plane is unknown before fits to data. To calculate the branching fraction, a weight was assigned to each event, $j$, as: 

\begin{equation}
\label{eq:sweight}
{\cal W}_j = \frac{\sum_iV_{sig,i}P_{i}(\vec{\alpha},\vec{x_j})}{\sum_kn_{k}{P_{k}(\vec{\alpha},\vec{x_j})}} ~,
\end{equation}

\noindent where $k$ is an integer and $V_{sig,i}$ is the signal row of the covariance matrix obtained from the fit~\cite{splot}. This procedure is effectively a background subtraction where these weights have the property $\sum_j{\cal W}_j = n_{sig}$. The branching fraction is then calculated as ${\cal B} = \sum_{j}{\cal W}_j/(\epsilon_{j} \times N_{\myBB})$  where $\epsilon_j$ (a function of $m_{K^{*+}h^-_1}^2$, $m_{h^-_1 h^+_2}^2$ and the $K^{*+} \to \Ks\pi^+$ decay helicity angle, ${\cal H}_{K^{*+}}$) varies across phase space and is simulated in bins using over seven million MC events for each channel. The size of the bins are optimized to provide continuous coverage of the efficiency distribution.

\begin{figure}[htb]
\resizebox{\columnwidth}{!}{
\includegraphics{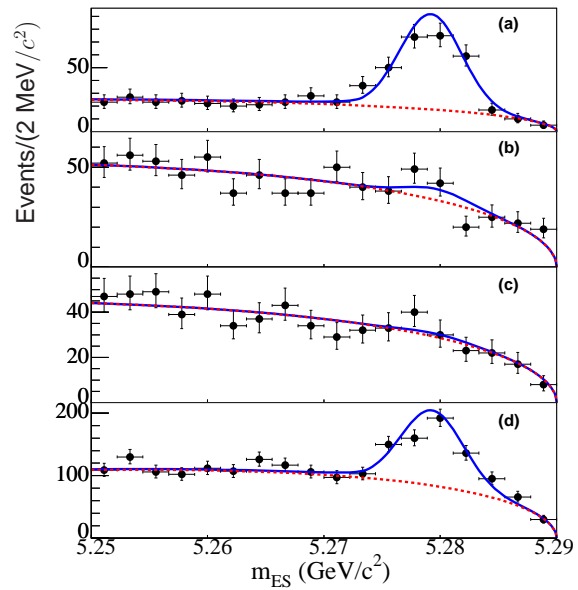}
}
\caption{Maximum likelihood fit projections of  $m_{ES}$ for signal-enhanced samples of charmless $B^+ \to K^{*+}h^+_1 h^-_2$ candidates. The dashed line is the fitted background PDF while the solid line is the sum of the signal and background PDFs. The points indicate the data. The plot labeled (a) shows a projection of $B^+ \to K^{*+}K^+ K^-$ candidates, (b) $B^+ \to K^{*+}\pi^+ K^-$ candidates, (c) $B^+ \to K^{*+}K^+ \pi^-$ candidates and (d) $B^+ \to K^{*+}\pi^+ \pi^-$ candidates.
}\label{fig:fitproj}
\end{figure}

Figure~\ref{fig:fitproj} shows the fitted projections for $B^+ \rightarrow K^{*+}K^+ K^-$, $B^+ \rightarrow K^{*+}\pi^+ K^-$, $B^+ \rightarrow K^{*+}K^+ \pi^-$ and $B^+ \rightarrow K^{*+}\pi^+ \pi^-$  candidates, while the fitted signal yield, measured branching fractions, upper limits  and asymmetries are shown in Table \ref{tab:results}. The candidates in Figure~\ref{fig:fitproj} are signal-enhanced, with a requirement on the probability ratio ${\cal P}_{sig}/({\cal P}_{sig} +{\cal P}_{bkg})$, optimized to enhance the visibility of potential signal, where ${\cal P}_{sig}$ and ${\cal P}_{bkg}$ are the signal and the total background probabilities, respectively (computed without using the variable plotted). The 90\% confidence level (C.L.) branching fraction upper limits (${\cal B}_{\rm UL}$) were determined by integrating the likelihood distribution (with systematics uncertainties included) as a function of the branching fraction from 0 to ${\cal B}_{\rm UL}$, so that $\int^{{\cal B}_{\rm UL}}_0 {\cal L}d{\cal B} = 0.9 \int^\infty_0 {\cal L}d{\cal B}$.

\begin{table*}[!ht]
\caption{Signal yields, efficiencies and branching fractions for $B^+ \rightarrow K^{*+}K^+ K^-$, $B^+ \rightarrow K^{*+}\pi^+ K^-$,  $B^+ \rightarrow K^{*+}K^+ \pi^-$ and $B^+ \rightarrow K^{*+}\pi^+ \pi^-$, measured using $\Ks\pi^+ h^+_1 h^-_2$ events. The first error is statistical and in the case of the measured branching fractions the second error is systematic. These efficiencies have taken into account that ${\cal B}(K^{*+} \to K^0\pi^+) = 2/3$, assuming isospin symmetry, as well as ${\cal B}(K^0 \to \Ks)$ and ${\cal B}(\Ks \to \pi^+\pi^-)$~\cite{pdg}. The branching fraction upper limits at a 90\% C.L. are shown for $B^+ \rightarrow K^{*+}\pi^+ K^-$ and  $B^+ \rightarrow K^{*+}K^+ \pi^-$. Asymmetries are reported (first error is statistical and the second is systematic) only for the channels with statistically significant yields.}\label{tab:results}
\begin{center}
\begin{tabular}{cccccc}
\hline
\hline
Mode&
Signal Events&
Efficiency&
Measured Branching Fraction &
Upper Limit&
Asymmetry (${\cal A}_{K^*hh}$)\\
&
Yield&
(\%)&
($\times$ 10$^{-6}$)&
($\times$ 10$^{-6}$)&
\\
\hline
$B^+ \rightarrow K^{*+}K^+ K^-$&
288 $\pm$ 26&
3.4&
36.2 $\pm$ 3.3 $\pm$ 3.6&
--&
0.11 $\pm$ 0.08 $\pm$ 0.03\\
$B^+ \rightarrow K^{*+}\pi^+ K^-$ &
20.1 $\pm$ 24.7&
3.5&
2.5 $\pm$ 3.1 $\pm$ 5.3 &
11.8&
--\\
$B^+ \rightarrow K^{*+}K^+ \pi^-$ &
9.7 $\pm$ 17.1&
3.5&
1.2 $\pm$ 2.1 $\pm$ 2.0 &
6.1&
--\\
$B^+ \rightarrow K^{*+}\pi^+ \pi^-$ &
583 $\pm$ 46&
3.3&
75.3 $\pm$ 6.0 $\pm$ 8.1&
--&
0.07 $\pm$ 0.07 $\pm$ 0.04\\
\hline
\hline
\end{tabular}
\end{center}
\end{table*}

Contributions to the branching fraction systematic error are shown in Table~\ref{tab:sys}. Errors due to charged tracking efficiency and $\Ks$ reconstruction efficiency were assigned by comparing control channels in MC events and data. The error in the efficiency was due to limited MC statistics, generated for each of the decays $B^+ \rightarrow K^{*+}K^+ K^-$, $B^+ \rightarrow K^{*+}\pi^+ K^-$, $B^+ \rightarrow K^{*+}K^+ \pi^-$ and $B^+ \rightarrow K^{*+}\pi^+ \pi^-$. Using a sample of $e^+e^- \rightarrow \mu^+\mu^-$ decays, the uncertainty in the number of \BB\ events was calculated to be 1.1\%. To calculate errors due to the fit procedure, a large number of MC samples were used, containing the amounts of signal and continuum events measured in data, and the estimated number of $B$-background events. The differences between the generated and fitted values using these samples were used to ascertain the sizes of any biases. Biases of +4.2, +10.7, +5.1 and +6.4\% were observed in the fitted signal yields of $B^+ \rightarrow K^{*+}K^+ K^-$, $B^+ \rightarrow K^{*+}\pi^+ K^-$, $B^+ \rightarrow K^{*+}K^+ \pi^-$ and $B^+ \rightarrow K^{*+}\pi^+ \pi^-$, respectively, that were a consequence of small correlations between fit variables. These biases were applied as corrections to obtain the final signal yields and half of the correction was added as a systematic uncertainty. The uncertainty of the $B$-background contribution to the fit was estimated by varying the known branching fractions within their errors. Each background was varied individually and the effect on the fitted signal yield was added in quadrature as a contribution to the uncertainty.  For the $m_{K^*}$ fit range there was also the possibility of $B$-background contributions from nonresonant and higher $K^{*+}$ resonances which was modeled in the data fit using a LASS parameterization~\cite{aston,latham}. The contribution from this background was estimated by extrapolating a $K\pi$ invariant mass projection fitted in a higher-mass region (0.992 $<$ $m_{K^*}$ $<$ 1.6 \gevcc), into the signal region. This estimated background was modeled in the final data fit, and half of the contribution to the signal was added as a systematic error contribution. The extrapolation assumed there were no integrated interference effects between the $K\pi$ background and the $K^{*+}(892)$ signal. The uncertainty due to reconstructing the  wrong $B^+ \rightarrow K^{*+}h^+_1 h^-_2$ signal candidate as a consequence of $K/\pi$ misidentification, for example $B^+ \rightarrow K^{*+}K^+ K^-$ events being reconstructed as $B^+ \rightarrow K^{*+}\pi^+ K^-$ candidates was determined using MC events and added as a systematic. The uncertainty due to PDF modeling was estimated from the calibration channel $B^+ \rightarrow \bar{D^0}(\rightarrow K^{*+} \pi^-) \pi^+$  and by varying the PDFs according to the precision of the parameters obtained from the calibration channel fit to data. In order to take correlations between parameters into account, the full correlation matrix was used when varying parameters. All PDF parameters that were originally fixed in the fit were then varied in turn, and each difference from the nominal fit was combined in quadrature and taken as a systematic contribution.  

\begin{table}[!ht]
\caption{Summary of systematic uncertainty contributions to the branching fraction measurements $B^+ \rightarrow K^{*+}h^+_1 h^-_2$. Multiplicative errors are shown as a percentage of the branching fraction and additive errors are shown in events. The final column shows the total systematic error on the branching fraction.}\label{tab:sys}
\begin{center}
\begin{tabular}{ccccc}
\hline
\hline
Error &
\footnotesize{$K^{*+}K^+ K^-$}&
\footnotesize{$K^{*+}\pi^+ K^-$}&
\footnotesize{$K^{*+}K^+ \pi^-$}&
\footnotesize{$K^{*+}\pi^+ \pi^-$}\\
source&
error(\%)&
error(\%)&
error(\%)&
error (\%)\\
\hline
\multicolumn{2}{c}{Multiplicative errors (\%)}&
&
 \\
Tracking&
2.4&
2.4&
2.4&
2.4\\
$\Ks$ Efficiency&
1.2&
1.2&
1.2&
1.2\\
Efficiency&
5.3&
9.4&
8.2&
5.6\\
No. of \BB\ &
1.1&
1.1&
1.1&
1.1\\
\hline
Tot. mult.(\%) &
6.0&
9.8&
8.7&
6.3\\
\hline
\multicolumn{2}{c}{Additive errors (events)}
&
&
 \\
Fit Bias&
6.0&
1.1&
0.3&
18.7\\
$B$-background&
6.9&
33.4&
1.3&
37.9\\
$K(892)^{*+}$ bkg&
19.5&
12.0&
4.5&
25.6\\
Signal mis-id&
0.0&
14.2&
15.5&
0.0\\
PDF params.&
8.4&
19.4&
3.0&
14.0\\
\hline
Tot. add.&
\raisebox{-1.5ex}[0cm][0cm]{23.1}&
\raisebox{-1.5ex}[0cm][0cm]{42.9}&
\raisebox{-1.5ex}[0cm][0cm]{16.5}&
\raisebox{-1.5ex}[0cm][0cm]{51.4}\\
(events) &
&
&
\\
\hline
\hline
Total (10$^{-6}$)&
3.6&
5.3&
2.0&
8.1\\
\hline
\hline
\end{tabular}
\end{center}
\end{table}

For the decays $B^+ \rightarrow K^{*+}h^+_1 h^-_2$, interference effects between the $K^{*+}(892)$ and S-wave final states (nonresonant and $K^{*+}_0(1430)$) integrate to zero if the acceptance of the detector and analysis is uniform; the same is true of the interference between the $K^{*+}(892)$ and D-wave final states ($K^{*+}_2(1430)$). Studies of MC events showed the efficiency variations were small enough to make these interference effects insignificant.  The integrated interference between $K^{*+}(892)$ and other P-wave amplitudes such as $K^{*+}_1(1410)$ is in principle nonzero, but in practice is negligible due to the small branching fraction of $K^{*+}_1(1410$) $\to \Ks\pi^+$ (6.6 $\pm$ 1.3\%~\cite{pdg}) and the fact that the $K\pi$ mass lineshapes have little overlap.

The $C\!P$-violating charge asymmetries for the decays $B^+ \rightarrow K^{*+}K^+ K^-$ and $B^+ \rightarrow K^{*+}\pi^+ \pi^-$ were measured to be ${\cal A}_{K^*KK} = 0.11 \pm 0.08 \pm 0.03$ and ${\cal A}_{K^*\pi\pi} = 0.07 \pm 0.07 \pm 0.04$, respectively, where the first errors are statistical and the second errors are systematic. The background asymmetries ${\cal A}^{Bkg}_{K^*KK}$ and ${\cal A}^{Bkg}_{K^*\pi\pi}$, which were expected to be consistent with zero, were measured to be 0.00 $\pm$ 0.02 and 0.01 $\pm$ 0.01, respectively. As a further study, the asymmetry ${\cal A}_{D\pi}$ for $B^+ \rightarrow \bar{D^0}(\rightarrow K^{*+} \pi^-) \pi^+$, also expected to be consistent with zero,  was measured to be $-0.01$ $\pm$ 0.02 with a corresponding background asymmetry ${\cal A}^{Bkg}_{D\pi}$ of $-0.01$ $\pm$ 0.06 (statistical errors only).

The systematic error on ${\cal A}_{K^*h_1h_2}$ was calculated by considering contributions due to track finding, fit biases, $B$-background uncertainties and particle interaction asymmetries. The error due to fit biases was found to be negligible. Uncertainties due to charged tracking efficiency were assigned by comparing control channels in MC simulation and data. The contribution from $B$-background was calculated by varying the number of expected events within errors and by assuming a $C\!P$-violating asymmetry of $\pm$ 0.2, as there are no available measurements for these decays. The $C\!P$ asymmetry assumed for the individual $B$-backgrounds was chosen to be greater than any asymmetry observed for charged $B$ decays. The interaction asymmetry of matter and antimatter with the detector was studied using background measurements ${\cal A}^{Bkg}_{K^*KK}$ and ${\cal A}^{Bkg}_{K^*\pi\pi}$, where no effect was observed and the statistical precision of the measurement was added as a systematic contribution.

In summary, we have analyzed $\Ks\pi^+ h^+_1 h^-_2$ final states to obtain the first branching fraction measurements for the decays $B^+ \rightarrow K^{*+}K^+ K^-$ and  $B^+ \rightarrow K^{*+}\pi^+ \pi^-$, and no evidence for new physics was found, placing upper limits on the branching fractions  $B^+ \rightarrow K^{*+}\pi^+ K^-$ and  $B^+ \rightarrow K^{*+}K^+ \pi^-$.

We are grateful for the excellent luminosity and machine conditions
provided by our \pep2\ colleagues, 
and for the substantial dedicated effort from
the computing organizations that support \babar.
The collaborating institutions wish to thank 
SLAC for its support and kind hospitality. 
This work is supported by
DOE
and NSF (USA),
NSERC (Canada),
IHEP (China),
CEA and
CNRS-IN2P3
(France),
BMBF and DFG
(Germany),
INFN (Italy),
FOM (The Netherlands),
NFR (Norway),
MIST (Russia), and
PPARC (United Kingdom). 
Individuals have received support from the
Marie Curie EIF (European Union) and
the A.~P.~Sloan Foundation.

\end{document}